# Faceted classification: management and use

Aida Slavic
*UDC Consortium, The Hague*

**Abstract:** The paper discusses issues related to the use of faceted classifications in an online environment. The author argues that knowledge organization systems can be fully utilized in information retrieval only if they are exposed and made available for machine processing. The experience with classification automation to date may be used to speed up and ease the conversion of existing faceted schemes or the creation of management tools for new systems. The author suggests that it is possible to agree on a set of functional requirements for supporting faceted classifications online that are equally relevant for the maintenance of classifications, the creation of classification indexing tools, or the management of classifications in an authority file. It is suggested that a set of requirements for analytico-synthetic classifications may be put forward to improve standards for the use and exchange of knowledge organization systems.

## 1. Introduction

It is generally acknowledged that subject access to information based on faceted classification offers more powerful and flexible information browsing and searching, and that this approach is particularly well suited for resource discovery on the Web. Topical hierarchical information organization, typical for resource organization on the Internet in the 1990s, for instance, has been repeatedly challenged by alternative and more flexible approaches.[1] What followed was a new 'take-up' on faceted classification that inspired applications, practical guidelines, machine readable formats and tools for building and implementing facet based interfaces. Another interesting 'twist' was the promotion of facet analysis as a more advanced way of knowledge organization and information design and architecture in general. Thus we find that the facet approach can be utilized within database design (Kayshap, 2003), user interface design (Allen, 1995), or for modelling a complex information space (Crystal, 2007). It comes as no surprise that we have also witnessed an increased interest in documentary faceted classification in recent years. Information specialists from various domains are tempted to study faceted schemes looking for a model, method or good practice examples of knowledge organization that can be emulated by new vocabularies.

In the past, once a scheme was published, the worry and cost of classification implementation online and its utilization in information retrieval was typically left to information services. As a result, significant resources were used in automating the same vocabulary, repeatedly over decades. Scores of researchers were solving the problem of classification searching and browsing at the user's end by devising programs to manage, decompose and structure the text of classification notation. In addition, countless projects worldwide duplicated effort in digitizing the same classification schedules and re-constructing them for machine processing.

Today, most indexing specialists would agree that the value of classification systems and vocabularies in general is measured by the ease and cost-efficiency of their implementation. It is considered that the requirements for classification use should be anticipated and embedded in schedules at source and that classification schemes should be exposed and/or distributed in a format that is more in line with the way they are used and the functions they have to support.

---

[1] Some general observation on this issue can be found in, for example, Ingwersen and Wormell (1992), Ellis &Vasconcellos (2002), Broughton and Lane (2004). Practical applications are described by Devadason (2003), Yee et al. (2003), Denton (2003), Van Dijk (2003).

A postulate of fundamental facet categories on which a faceted classification system may be organized prescribes the rule of ordering of classes and governs analytico-synthetic functions (citation order). The most important characteristic of an analytico-synthetic classification is that it contains composite notational expressions based on facet organization that can be decomposed into semantically meaningful elements. From this, it transpires that one of the most obvious functions that has to be supported is access to and management of facets in the schedules and within composite notational elements.

In this paper we argue that there are concerns in classification design, implementation and use that ought to be anticipated at the point of classification creation; i.e. at the very source. For, irrespective of the quality of their intellectual content or robust logic underlying their structure, faceted classification schemes have little or no use in information retrieval if their structural and functional characteristics are not exposed. By focusing on some technical details we would like to illustrate how it may be possible to find an area in which some patterns in faceted classification design and formatting classification for machine processing can be standardized. We suggest that it should be possible to discern some functional requirements common to analytico-synthetic classifications and those ought to find their place in vocabulary standards.

## 2. Classification automation in bibliographic domain

During the 1970s and 1980s classification indexes were made available for searching and browsing in online public access catalogues (OPACs). As early as 1983, Wajenberg observed a discrepancy between the composite structure and dense classification notation semantics, and their primitive processing in library systems that impeded classification searching. He, for instance, recommended extending MAchine Readable Cataloguing (MARC) codes for Dewey Decimal Classifications (DDC) so as to exploit shelfmarks for improving subject searching of bibliographic data.

At the time of Wajenberg's proposal there were no library classifications held, maintained or distributed in an electronic form. The research by Cochrane and Markey (1985) on DDC was the first step in that direction. They provided an exhaustive list of classification data elements needed to support searching and browsing. This effectively represented a subject authority control framework that was independent of the bibliographic description and was capable of serving classification maintenance, distribution and its subsequent use in information retrieval.

Although, by the 1990s, librarians fully embraced authority control for managing names, it took some time for the same model to be proposed as a solution for supporting subject data, including classification. Mandel (1995), Barnhart, (1996), Guenther (1996) and Cochrane and Johnson (1996) all emphasised the obvious fact that access points to classification data ought to be controlled centrally and independently of bibliographic records.

Independent of the trends in the abovementioned automation of library systems, DDC and UDC owners and publishers created database tools primarily to facilitate the maintenance and publishing of classifications. For these tools the following functions were deemed essential:

- searching and browsing of classification notation
- searching notation through associated verbal expression

- sort and display
- automatic tracing of hierarchical and associative linking
- tracing of system rules to the area of their application
- navigation between tables, facets and subject areas
- tracing historical data through a scheme's lifespan ('replaces/replaced by')
- various outputs and exports
- identification of classes independent of notation

These database tools were for classification editors only and were not very user friendly with respect to searching and browsing. The issue of classification data being able to support user friendly hierarchical browsing, semantic linking and facet control only became central to classification owners when publishers started selling access to schedules on CD ROMs (from 1995 onwards), and on the Web (from 2001 onwards). The contribution to the automation of classification from the publishers of DDC and UDC is significant, for when a classification is converted into a machine readable format the same data can be used to support various tools for machine assisted indexing or authority control tools for supporting information retrieval (cf. Markey, 2006).

In 1992 we had the first important 'breakthrough' with respect to classification use online when the *USMARC Format for Classification Data* was created as an independent format to meet the needs of expressing classification data. The creation of this format was to some extent driven by the plans for the conversion of the Library of Congress Classification (LCC), to which purpose it was effectively used in 1993 (Guenther, 1996). This standard was later superseded by *MARC 21 Concise Format for Classification Data* in 2003 which has since been updated and improved through subsequent versions. Although in theory the format was presented as a general solution, in practice it was based on LCC and DDC structures and is for this reason better suited for enumerative classifications (Slavic and Cordeiro, 2004b). The next opportunity to 'rethink' and improve machine readability was occasioned with work on the UNIMARC authority format for classification data that was meant to be improved to support analytico-synthetic systems. Unfortunately, proposals for structuring the classification number field to reflect composite notations were declined as 'unnecessary complication' and the *Concise UNIMARC Classification Format* was published in its draft form in 2000, completely mirroring the earlier MARC 21 format. This format is still awaiting official completion. MARC classification formats are meant to resolve what was often felt as the weakest points of library OPACs: subject browsing and retrieval. It is unfortunate that neither of them is entirely suited to managing the composite notation typical of faceted systems.

The level of granularity of the coding of data elements in the above formats is driven by the functionality that the format has to provide, and logically each classification element whose manipulation will support searching, maintenance or display is coded as a separate element. The most important function of the format is the successful separation of classification data content and structure from the way data are displayed in print or in an online system.

### 3. Separating content, presentation and function

Typically a class in a documentary classification schedule would be represented in the following manner (example from UDC):

| 004.421.2 | Basic mathematical algorithms |
|---|---|
| | *For mathematical theory of algorithms in general use 510.5* |
| | *Specify mathematical process by colon combination with class 51* |

*Examples of combination(s)*
004.421.2:517.443 Fast Fourier transform
004.421.2:517.535 Algorithms for rational expression
004.421.2:519.17 Graph algorithms
⇒ *519.16*
⇒ *519.178*

When stored in a database, information implicit in the class information showed above will have to be made explicit using following 7 blocks of data elements:

1. **NOTATION (CLASSIFICATION NUMBER)**

    tables from which notation is taken
    type of notation (simple or composed)
    notation structural elements/components
    relationships between elements: span, phase relationships

2. **BROADER CLASS**

3. **CAPTION**

4. **NOTES**

    SCOPE NOTE
    APPLICATION (INSTRUCTION) NOTE
    NOTATION BUILDING NOTES AND RULES

    Rules for parallel division [2](derived from; divide)
    Rules for combination and expansion (add, specify by)
    Examples of combination
    NOTATION HISTORY NOTE (replaces, replaced by)
    GENERAL CONTENT NOTE
    EDITORIAL NOTE

5. **REFERENCES (SEE ALSO)**

6. **CLASS ID** (unique identifier of a class)

7. **INDEX (SEARCH) TERMS** (keywords)

We can think of these blocks of data as a standard container that we have available to record more detailed information from a specific system.

If analysing analytico-synthetic systems such as Colon Classification (CC), Bliss Bibliographic Classification (BC2), Broad System of Ordering (BSO) and UDC, we notice that they all share the following characteristics:

- a system as a whole is organized into disciplines and sub-disciplines (main tables or main schedules)
- generally applicable concepts (time, place, languages, materials, form etc.) are kept in separate tables and can be attached to any notation from the main table
- disciplines and sub-disciplines are structured into facets/sub-facets - classes from different facets can be combined to express a compound subject
- there are rules for ordering facets in a complex notational expression (citation order) and for ordering of classes in the schedules.

---

[2] Parallel division rule by which the same subdivision can be used in several places in the schedules. (e.g. animals in zoology, palaeontology, agriculture; languages in ethnic grouping, literature and linguistics ). So, for instance, UDC notation *821.**111** English literature* or *(=**111**) English speaking people* is derived from a notation *=111 English language.*

- subjects from different disciplines can combined using relators or symbols for phase relationships (influence, comparison, application, bias)

(cf. Slavic and Cordeiro, 2004a; Gnoli, 2007)

A classification management tool supporting these kinds of systems needs to control the following:

- logical ordering (sorting) of classes
- syntactic relationships
    citation order (the order of facets in a pre-coordinated classmark)
    phase relationships (relationship between two or more independent subjects)
- hierarchical relationships
- associative linking (see also)

Information necessary to support these functions can be recorded in one or all of the following ways: a notational system; schedule presentation; and machine readable coding.

## 4. The role of notation in classification use online

Classification notation is the part of a classification system that is added to schedules upon the completion of classification structure as a complementary or subsidiary element. Usually defined as a system of symbols representing classification structure, notation's main function is the mechanical arrangement of classes. As it may be used for book shelf arrangement, notation brevity is an important issue for classificationists (Ranganathan, 1962; Vickery, 1952-1959); Broughton, 1999).

But an important function of notation is also to display the structure and reveal or suggest relations between classes. Hence we speak of 'notational expressiveness'. This function of notation is very important when converting schedules into a machine readable format and can be utilized to underpin classification management and use online.

In order to support compound formation, the notation in an analytico-synthetic scheme is structured of the following parts: notation for terms (isolate and composite), facet indicators (signposts for facets) and phase relations symbols (indicating relationships between subjects). For obvious reasons, the notation of an analytico-synthetic system tends to be longer and more complex. To address this 'problem' some classifications use techniques for shortening the notation at the expense of expressiveness: telescoping of arrays or retroactive notation building (Vickery, 1952-1959; Broughton, 1999). Thus in some analytico-synthetic classifications we may have composite notations produced by simple juxtaposition (without connecting symbols), or even by dropping the initial digits resulting in a notation which Vickery termed 'amalgamated' (cf. Slavic and Cordeiro, 2004b; Gnoli, 2007).

As a result we ought to be prepared to deal with expressive and non-expressive notation (Broughton, 1999). When expressive, the notation may be so: a) with respect to hierarchy, having decimal form and b) with respect to syntax, having facet indicators and phase relators. Equally, notation may appear completely 'un-informative' with respect to hierarchy. When this is the case, hierarchy has to be expressed using 'indentation' which then itself becomes a representation of logical division that ought to be captured and translated into machine readable instructions.

| Hierarchically and syntactically expressive notation | | Non-expressive notation | |
|---|---|---|---|
| -A | Africa | A | Africa |
| -A5 | Central and eastern Africa | AZ | Central and eastern Africa |
| -A51 | Uganda | AZA | Uganda |
| -A511 | Northern region | AZB | Northern region |
| **-A5111** | Yumbe | **AZC** | Yumbe |
| Z | Threats to the environment | Z | Threats to the environment |
| Z1 | Threat of depredation | ZB | Threat of depredation |
| Z13 | Depletion by over exploitation | ZC | Depletion by over exploitation |
| **Z132** | Deforestation | ZF | Deforestation |
| Z132A5111 | | ZFAZC | |

As shown in the example above the most important characteristic of syntactically expressive notation is that each structural component is 'declared' by a specific beginning or facet indicator expressed by some kind of symbol indicating the facet from which the term is taken.

The importance of facet indicators in managing classification was emphasised by authors concerned with information retrieval functionalities of faceted systems such as Gödert (1991), Gopinath and Prasad (1994), Pollitt and Tinker (2000), Madalli and Prasad (2002), Gnoli and Hong (2006). The authors developing programs for automatic decomposition of classification numbers in library catalogues, such as Liu (1996) and Riesthuis (1998), have also reported on the advantages of expressive notation when defining algorithms for 'splitting' composite notations in DDC and UDC respectively.

Expressive notation may also be a useful tool to lean on when designing new classifications - for it will be easy way of recording structural rules and imposing rigidity in data representation. In a database this can also be utilized as complementary tool for control and validation or in creating a machine assisted indexing tool. As demonstrated in some recently built faceted classification systems and tools - developing an expressive notational system may be well suited for indexing digital resources for which the length of notation is not viewed as a disadvantage (Gnoli and Hong, 2006; Broughton and Slavic, 2007). In this context notation need not to be handled by humans and need not to be displayed or searched.

When choosing or considering a notational system we have to take into consideration the way classification is applied online. Building classification is a long intellectual process that has always been and will continue to be an isolated task, but when devising notation the authors may be advised to anticipate representational consistency and rigour that will be required later on by programs.

When a class is composed it has to be represented as such and the easiest way to do so is through an expressive notation. This will facilitate multidirectional access to every semantically significant element of a synthesized notation on which the following functions may be built:

- coordinate searching of notation
- link notation to verbal expressions for the purpose of searching using words
- support global changes in a classification authority tool in case of reclassifications or cancellations
- controlling display (e.g. altering the presentation of parts of notation, colouring, replacing by other symbols)
- controlling citation order and sorting order in an indexing tool

- supporting facets browsing/view-based browsing

Once all the elements of the scheme structure are machine readable, display and use of classification online becomes independent of notation. Theoretically one can 'detach' a notational system and replace it or link it to a parallel system; for example, if shorter classmarks are needed for book labels in a physical collection. If the system is supported by a proper indexing tool - indexers may never need to key in classification numbers or build and re-build synthesised classmarks manually. With a help of a classification authority file, for instance, once a composed notation expression is recorded it becomes reusable, saving labour and ensuring consistency. In summary it could be said that expressive notation may help and is not likely to impose any problems in classification use online.

## 5. Managing logical order of classes in an online environment[3]

Presentation of a knowledge area in any documentary classifications is from general to specific and from abstract to concrete. To achieve this, analytico-synthetic classifications have to control the sequence of concepts in both the hierarchical order of concepts (paradigmatic relationship) and in the syntactic order of concepts (syntagmatic relationship).
Facet classifications achieve this by organizing concepts into fundamental facet categories that follow a particular sequence. For instance, the fundamental facets in Bliss Bibliographic Classification follow the citation order "*Thing-Kind-Part-Property-Material-Process-Operation-Patient-Product-Byproduct-Agent-Place-Time*" (from concrete to abstract or from specific to general). Thus when 'building' a notation by adding concepts from various facets, guided by this citation formula, we will state the most concrete/specific concept 'Thing' as first and will continue in the direction of decreased concreteness/specificity. A field of knowledge, however, is presented following a general to specific order. Hence the order of facets in the schedule will be the opposite of the order in which these facets are listed in composite notation.

The order can be managed through notation by labelling each facet in such a way that the required order in the schedule will be achieved mechanically. The citation order, however, will be governed by a rule of 'reversed schedule order' and the building of numbers will have to respect this sequence:

---

Order of facets in schedules achieved mechanically (alphabetical order of facet indicators):

---

*Subject area*          **33**

Facets in 33
    *Time*               A1      ← most general/abstract
*Space*                 B1
*Agent*                 C1
Product                 D1
...
*Thing*                 M1      ← most specific/ concrete

---

[3] The ordering in faceted classifications as discussed here is significantly simplified - and it 'flags' only the most frequent and obvious problems. For more complex problems of ordering that result in unhelpful and illogical sequences, see Vickery (1956), Broughton (1999), Gnoli (2007).

| |
|---|
| Citation order in a synthesised classmark - facets cited in reversed alphabetical order: |

**33**M1*D1C1B1A1*
specific/concrete → general/abstract

Following this simple technique, composed notations sorted mechanically will always produce a general-to-specific sequence of subjects:

> 33A1
> 33B1*A1*
> 33D1
> 33M1
> 33M1*B1*A1
> 33M1*C1*B1*A1*
> 33M1*D1*
> 33M1*D1*C1
> 33M1*D1*C1*B1*A1

Following this simple technique, when facets are coded for processing, the logical ordering established by schedules can be automatically mirrored in the database formats.

Ordering will require special attention when involving composite notations expressing phase relations between two subjects. Here is an example of a different kind of relationship, taken from the FAT-HUM classification (cf. Broughton and Slavic, 2007):

|   |   |   | Example |
|---|---|---|---|
| Addition | + | 590+420 | 'education in addition to religion' |
| Range | / | 420/590 | 'the field of subjects spanning education to religion' |
| Coordination | : | 590:420 | 'religion and education in reciprocal relationships' |
| Comparison phase | = | 590=420 | 'comparison between religion and education' |
| Influence phase | >> | 590>>420 | 'the influence of religion on education' |
| Bias phase | << | 420<<590 | 'religion for educational purposes' |
| Exposition phase | - | 590-420 | 'religion as viewed by education' |
| Sub-grouping | < | 590<420 | 'education as part of religion' |

In these cases the order of classes is not mechanical and has to be specified by the scheme designers. Based on instructions provided, 'relationship symbols' will be assigned codes that impose a desired order in the machine readable expression of the scheme.

5.1     Correcting the hierarchy

Probably the most often cited advantages of classification used in retrieval, in the 1980s and 1990s, was that classifications with expressive notation support semantic search expansion. In most cases, by removing the last digit of the notation we will automatically retrieve documents in a broader class. However, 'most' may not be good enough. Another misconception of 'ease' comes from a browsing interface that allows expansion and collapsing of the hierarchy. In spite of the natural feel, this functionality should not be taken for granted. In order to process the hierarchy, programs need information coming from the coding (tracing) of the relationship between each notation and its immediate broader class. In theory programs can be written to do this automatically from the hierarchically expressive notation. In real life, there are a couple of situations where manual correction is required.

Firstly, special attention will be required when we have relations between classes. The most common example across classification systems, for instance, are classes represented as a range of subsequent classes, represented with a range symbol, usually / (forward stroke). For instance, A1/A2 represents the broader class of A1 and A2. However, when concept A1 relates to another concept e.g. A1:A2 this represents a narrower, more specific subject which is logically subsumed to class A1. Hence the correct hierarchy would be:

   A1/A2
    A1
     A1:A2
    A2
    A3
    A4

If these situations are properly recorded, the coding in the management tool will control the appropriate ordering. Manual corrections will, however, be required, for various cases of 'invalid' hierarchies listed below.

### 5.1.1 Telescoped arrays

Classes that are subordinate are presented in notation as coordinated. This is done on purpose in order to reduce the length of notation. This is usually in accordance to Ranganathan (1962) rules for telescoping arrays in library classification[4] which he has introduced as an exception to his *canon of hierarchy*, for example:

| | |
|---|---|
| (1) | Place and space in general |
| | [array 2 'telescoped' to array 1] |
| (3) | Ancient world |
| (4) | Europe |
| (5) | Asia |
| (6) | Africa |
| (7) | North and Central America |
| (8) | South America |
| (9) | South Pacific and Australia |

### 5.1.2 Missing levels

This occurs (a) with decimal notations when class subdivision is centesimal to accommodate more than ten subdivisions, (b) when superfluous classes are removed, (c) when a very specific composite notation is added to a class while immediate super-ordinate classes are not introduced, for example:

| | |
|---|---|
| 271 | Eastern Church |
| 271.2 | Orthodox Church |
| 271.2-282.7 | Service books - Orthodox Church [missing] |
| 271.2-282.7-2 | Service books - Evidence of religion - Orthodox Church [missing] |
| 271.2-282.7-24 | Service books - Specific texts - Orthodox Church [missing] |
| 271.2-282.7-247 | The Gospel Book |

---

[4] Definition of a telescoped array: "*Array of classes in a schedule of classification, made of co-ordinate and subordinate isolates, as viewed from the Idea Plane, but whose class numbers appear to be co-ordinate, as viewed from Notation Plane*" (Ranganathan, 1962: 278).

*5.1.3    'False' hierarchy*

It occurs when classes appear subordinated when in fact they are not. This may be found occasionally in older classification schemes when old and new subdivisions are merged (often in area tables). In the following example, Galicia, which is the name for the former territories that are now shared between southern Poland and western Ukraine, appears in this UDC notation to be part of the former Czechoslovakia. Bukowina, which was the southeast part of Galicia, does not appear to be part of Galicia but rather a part of Czechoslovakia when, in fact, at present this is the land between Romania and the Ukraine.

| | |
|---|---|
| (437) | Czechoslovakia (1918-1992) |
| (437.3) | Czech Republic |
| **(437.4)** | **Galicia (Galizien) to 1919** |
| **(437.5)** | **Bukowina to 1919** |
| (437.6) | Slovakia. Slovak Republic. Slovenská Republika |
| (437.7) | Zakarpatska Ukrajina (1918-1938) |

The above cases are typical for classification schemes that have been in use for a long time. However, even new schemes that are being created as we speak may not be protected from various notational defects. Because of the inevitability of 'faults' in a notational hierarchy, the basic requirement in management and maintenance is to be able to manually control and correct data on a 'broader class' link.

**6.   Standards for vocabulary exchange and faceted classifications**

Vocabulary standards are vehicles for vocabulary implementation and sharing. Until recently we have been orientated on domain specific standards such as MARC but knowledge organization systems are now used and exchanged beyond the bibliographic domain. Recent development of networked standards for knowledge organization systems; primarily ISO/IEC 13250 Topic Maps (2000), BS 8723 Structured vocabularies for information retrieval (2005) and Simple Knowledge Organization System (SKOS) (2006), have been created to support sharing and use of vocabularies across systems and domains. The change of name, scope and objectives of BS8723 (2005-), which is a standard replacing the old *Guidelines for thesaurus construction,* is indicative of this trend. The new BS8723 standard will not only cover all types of vocabularies expressing hierarchical relationships but will also provide a standard XML schema for their export. As a general rule all of the abovementioned standards offer XML encoding schemes as platform and system neutral carriers for transporting data.[5]

The general idea behind these standards is that the availability and exchange of controlled vocabularies in an open networked environment may contribute to resource discovery directly or indirectly through referencing, resolving language ambiguities and providing semantic context for text processing. With vocabularies expressed in a standardized way, there is a realistic prospect of centrally managed vocabulary repositories and services that will facilitate cross-collection resource discovery through vocabulary mappings and translation (Tudhope et al. 2006).

---

[5] We can mention here a simple format with XML encoding created to express resource description terms organized in mutually exclusive categories (facets) under the title XFML (eXchangeable Faceted Metadata Language) - which does not have the status of a vocabulary standard but is frequently used to support applications for facet-based Web interfaces (Van Dijk, 2003). This schema also seems to be focused on hierarchical relationships and has no provision for accommodating complex vocabularies.

Topic Maps, for instance, offer a framework that can be used to accommodate machine understandable representations of subject vocabularies. SKOS and BS8723 strive towards practical vocabulary encoding solutions. At the moment, however, none of these standards address the requirements typical of pre-coordinated indexing languages and have no solution for expressing syntagmatic relationships. If using the available standards, the co-ordinate relationships between concepts in an analytico-synthetic classification would not be expressed. This issue is yet to be addressed through collaborative work between classificationists and working groups within specific standards.

## 7. Conclusion

When a classification is available in a machine readable format its potential for indexing and information retrieval can be fully exploited. Faceted classifications, being structurally and syntactically more complex, are even more likely to benefit from machine processing. Once available the same data can easily be utilized to support: tools for classification management and distribution; tools for machine assisted indexing; and authority control underpinning information retrieval. In addition, if exposed in a standard format for vocabulary exchange, faceted classifications would become more accessible to people and systems.

In this paper we have attempted to create an awareness of some issues in data presentation and some general questions of 'translating' the structural components of a classification scheme into machine readable expressions. Our concern was primarily with documentary classifications with an analytico-syntactic structure which tend to be more complex and whose structural rules may not always be as clearly declared or obvious. Our main goal was to illustrate the amount of detail involved in bringing faceted classification 'alive' and to suggest that none of the advancements in classification browsing and searching can be taken for granted. Our other intention was to encourage classification designers, even when not involved in management tool design, to record more and more rigorous data about classification structure from the outset. The sooner we start to think about some neutral and objective way of representing a specific classification scheme, the sooner we will advance towards some common models for faceted systems that may find their way into vocabulary standards.

It is common sense that in the future, creators and publishers expose their schemes in such a format that will facilitate their use, thus reducing the duplication of work and implementation costs. This task would be more feasible if existing vocabulary standards would be able to express classifications with a typified analytico-synthetic structure. It is, therefore, very important that classification developers, owners and users alike become more involved in standards development.